\documentclass[a4paper,12pt,titlepage]{article}

\usepackage{amsmath,amssymb,revsymb}
\usepackage[dvipdfmx]{graphicx}
\usepackage{here}
\usepackage{bm}
\usepackage{fancybox}
\usepackage{secdot}
\usepackage{comment}
\usepackage{braket}
\usepackage{titlesec}
\usepackage{geometry}
\usepackage{hyperref}
\usepackage{slashed}
\usepackage{fancyhdr}
\usepackage{boxedminipage}
\usepackage[dvipdfmx]{color}
\usepackage{mathrsfs}
\usepackage{feynmf}
\usepackage{cite}

\titleformat*{\section}{\LARGE\bfseries}
\titleformat*{\subsection}{\Large\bfseries}

\hypersetup{
setpagesize=false,
 bookmarksnumbered=true,%
 bookmarksopen=true,%
 colorlinks=true,%
 linkcolor=blue,
 citecolor=blue,
 }
 
\makeatletter 
\long\def\@makefntext#1{\parindent 1em\noindent 
\@hangfrom{\hbox to 1.8em{\hss$^{\@thefnmark}$}}#1}
\makeatother

\begin{document}

\fancypagestyle{foot}
{
\fancyfoot[L]{$^{*}$E-mail address : yakkuru$\_$111@ruri.waseda.jp\\
$^{\dagger}$E-mail address : msato@hirosaki-u.ac.jp}
\fancyfoot[C]{}
\fancyfoot[R]{}
\renewcommand{\headrulewidth}{0pt}
\renewcommand{\footrulewidth}{0.5pt}
}

\renewcommand{\footnoterule}{%
  \kern -3pt
  \hrule width \columnwidth
  \kern 2.6pt}


\begin{titlepage}
\begin{flushright}
\begin{minipage}{0.2\linewidth}
\normalsize
WU-HEP-20-06
\end{minipage}
\end{flushright}

\begin{center}

\vspace*{5truemm}
\Large
\bigskip\bigskip

\LARGE\textbf{String Backgrounds in String Geometry}%

\Large

\bigskip\bigskip
Masaki Honda$^{1,*}$ and Matsuo Sato$^{2,\dagger}$
\vspace{1cm}

{\large$^{1}$ \it{Department of Physics, Waseda University, 
Totsuka-cho 1-104, Shinjuku-ku, Tokyo 169-8555, Japan}}\\
{\large$^{2}$ \it{Graduate School of Science and Technology, Hirosaki University, Bunkyo-cho 3, Hirosaki, Aomori 036-8561, Japan}}\\
\bigskip\bigskip\bigskip
\large\textbf{Abstract}\\
\end{center}
String geometry theory is a candidate of the non-perturbative formulation of string theory. In order to determine the string vacuum, we need to clarify how string backgrounds are described in string geometry theory. In this paper, we show that the string backgrounds are embedded in configurations of the fields of a string geometry model. Especially, we show that the configurations satisfy the equations of motion of the string geometry model if and only if the embedded string backgrounds satisfy their equations of motion. This means that classical dynamics of the string backgrounds are described as a part of classical dynamics in string geometry theory. Furthermore, we define an energy of the configurations in the string geometry model because they do not depend on the string geometry time. A string background can be determined by minimizing the energy.

\thispagestyle{foot} 

\end{titlepage}

\baselineskip 7.5mm


\tableofcontents


\parindent=20pt

\jot=8pt

\renewcommand\thefootnote{\textcolor{red}{\arabic{footnote}}}

\setlength{\abovedisplayskip}{12pt} 
\setlength{\belowdisplayskip}{12pt} 

\section{Introduction}
Superstring theory is a promising candidate of a unified theory including gravity. However, superstring theory is established at only the perturbative level as of this moment. The perturbative superstring theory lacks predictability because it has many perturbatively stable vacua. 

String geometry theory is a candidate of non-perturbative formulation of superstring theory~\cite{Sato:2017qhj}, which can determine a non-perturbatively stable vacuum. In string geometry theory, the path-integral of the perturbative superstring theory on the flat string background is derived by taking a Newtonian limit of fluctuations around a fixed flat background in an Einstein-Hilbert action coupled with any field on string manifolds~\cite{Sato:2017qhj, Sato:2020szq}\footnote{A perturbative topological string theory is also derived from the topological sector of string geometry theory \cite{Sato:2019cno}. }. That is, the spectrum and all order scattering amplitudes in superstring theory on a flat background are derived from string geometry theory. However, perturbative string theory describes only propagations and interactions of strings in a fixed classical string background, and cannot describe dynamics of the classical string background itself. Only the consistency with the Weyl invariance requires that the string background satisfies the equations of motion of supergravity. In order to determine a string background, a non-perturbative string theory needs to be able to describe dynamics of the string backgrounds not in consequence of consistency. Especially, configurations that represent the classical string backgrounds should satisfy the equations of motion of a non-perturbative string theory.

In this paper, as a first step to determine the string vacuum, we show that configurations that represent the classical string backgrounds satisfy the equations of motion of string geometry theory. First, in string geometry theory, we define configurations of fields that include the string backgrounds. Second, we show that the equations of motion where the configurations are substituted are equivalent to the equations of motions of the string backgrounds. That is, the action of the string geometry theory is consistently truncated to the action of the string background. This means that classical dynamics of the string backgrounds are described as a part of classical dynamics in string geometry theory. This fact supports the conjecture that string geometry theory is a non-perturbative formulation of string theory. Moreover, we study how a string background can be determined. We focus on the bosonic string background since our essential argument lies in the bosonic case. A supersymmetric extension will be given in \cite{Honda-Sato}.

The organization of this paper is as follows. In Sec.~2, we introduce a string geometry model. In Sec.~3, we identify string background configurations and obtain the equations of motion of the string background from the equations of motions of the string geometry model by a consistent truncation. In Sec.~4, we define an energy of the string background configurations because they do not depend on the string geometry time. A string background can be determined by minimizing the energy. In Sec.~5, we conclude and discuss our results.

\section{String geometry model}
We study a string geometry model\footnote{The action of string geometry theory is not determined as of this moment. On this stage, we should consider various possible actions. Then, we call each action a {\it string geometry model} and call the whole formulation {\it string geometry theory}. In \cite{Sato:2017qhj}, the perturbative string theory on the flat spacetime is derived from a gravitational model coupled with a $u(1)$ field on a Riemannian string manifold, whereas in \cite{Sato:2020szq}, it is derived from gravitational models coupled with arbitrary fields on a  Riemannian string manifold. Thus, the perturbative string theory on the flat spacetime is derived from this model.} whose action is given by,
\begin{align}
\label{action of bos string-geometric model}
&S = \frac{1}{G_{N}} \int  \mathcal{D} h \mathcal{D} X_{\hat{D}_{T}}(\bar{\tau})   \mathcal{D} \bar{\tau} \sqrt{ \mathbf{G} } e^{-2 \Phi} \left[ \mathbf{R}  + 4 \nabla_{I} \Phi \nabla^{I} \Phi - \frac{1}{2} |\mathbf{H} |^{2}   \right], 
\end{align}
where $G_{N}$ is a constant, $I=\{d,(\mu \bar{\sigma}) \}$, $| \mathbf{H} |^{2}:= \frac{1}{3!} \mathbf{G}^{I_{1} J_{1}} \mathbf{G}^{I_{2} J_{2}} \mathbf{G}^{I_{3} J_{3}} \mathbf{H}_{I_{1} I_{2} I_{3}} \mathbf{H}_{J_{1} J_{2} J_{3}}$, and we use the Einstein notation for the index $I$. The action~(\ref{action of bos string-geometric model}) consists of a metric $\mathbf{G}_{I_{1} I_{2}}$, a scalar field $\Phi$ and field strengths $\mathbf{H}_{ I_{1} I_{2} I_{3} }$ of a two-form field $\mathbf{B}_{I_{1} I_{2}}$. They are defined on a {\it Riemannian string manifold}, whose definition is given in \cite{Sato:2017qhj}, which we summarize here. It is an infinite dimensional Riemannian manifold parametrized by coordinates $(\bar{h},X_{\hat{D}_{T}}(\bar{\tau}), \bar{\tau})$\footnote{$\,\, \bar{}$ represents a representative of the diffeomorphism and Weyl transformation on the worldsheet. Giving a Riemann surface $\bar{\Sigma}$ is equivalent to giving a  metric $\bar{h}$ up to diffeomorphism and Weyl transformations.}, where $\bar{h}$ is the metric on a worldsheet $\bar{\Sigma}$\footnote{$\bar{h}$ is a discrete variable in the topology of string geometry, where an $\epsilon$-open neighbourhood of $[\bar{h}, X_s(\bar{\tau}_s), \bar{\tau}_s]$ is defined by
\begin{eqnarray}
U([\bar{h}, X_s(\bar{\tau}_s), \bar{\tau}_s], \epsilon)
:=
\left\{[\bar{h},  X(\bar{\tau}), \bar{\tau}] \bigm| \sqrt{|\bar{\tau}-\bar{\tau}_s|^2+  \| X(\bar{\tau})-X_s(\bar{\tau}_s) \|^2} <\epsilon   \right\}.
\label{neighbour}
\end{eqnarray}
As a result, $d \bar{h}$  cannot be a part of basis that span the cotangent space in (\ref{cotangen}), whereas fields are functionals of $\bar{h}$ as in (\ref{LineElement}).}, 
$\bar{\tau}$ is the global time on $\bar{\Sigma}$ and $X_{\hat{D}_{T}}(\bar{\tau})$ is a map from $\Sigma|_{\bar{\tau}}$ to the 26-dimensional Euclidean space $\mathbb{R}^{26}$. $\hat{D}_T$ represents all the backgrounds except for the target metric, that consist of the B-field and the dilaton. The cotangent space is spanned by 
\begin{eqnarray}
d X^{d}_{\hat{D}_{T}} &:=& d \bar{\tau}
\nonumber \\
d X^{(\mu \bar{\sigma}) }_{\hat{D}_{T}}&:=& d X^{\mu}_{\hat{D}_{T}} \left( \bar{\sigma}, \bar{\tau} \right), \label{cotangen}
\end{eqnarray}
where $\mu=1, \dots, 26$. 
The summation over $\bar{\sigma}$ is defined by 
$\int d\bar{\sigma} \bar{e}(\bar{\sigma},\bar{\tau})$,
where
$\bar{e}(\bar{\sigma},\bar{\tau}):= \sqrt{\bar{h}_{\bar{\sigma} \bar{\sigma}}}$. This summation is transformed as a scalar under $\bar{\tau} \mapsto \bar{\tau}'(\bar{\tau}, X_{\hat{D}_T}(\bar{\tau}))$ and invariant under $\bar{\sigma} \mapsto \bar{\sigma}'(\bar{\sigma})$.
For example, an explicit form of the line element is given by
\begin{align}
&ds^{2} (\bar{h}, X_{\hat{D}_{T}}(\bar{\tau}), \bar{\tau}) \nonumber \\
= &\mathbf{G}_{dd}(\bar{h}, X_{\hat{D}_{T}}(\bar{\tau}), \bar{\tau})(d \bar{\tau} )^{2} + 2 d \bar{\tau} \int d \bar{\sigma} \bar{e}(\bar{\sigma},\bar{\tau}) \sum_{\mu} \mathbf{G}_{d (\mu \bar{\sigma})}(\bar{h}, X_{\hat{D}_{T}}(\bar{\tau}), \bar{\tau}) d X^{\mu}(\bar{\sigma},\bar{\tau}) \nonumber \\
&+ \int d \bar{\sigma} \bar{e}(\bar{\sigma},\bar{\tau}) \int d \bar{\sigma}' \bar{e}(\bar{\sigma}',\bar{\tau}) \sum_{\mu,\mu'} \mathbf{G}_{(\mu \bar{\sigma})(\mu' \bar{\sigma}')}(\bar{h}, X_{\hat{D}_{T}}(\bar{\tau}), \bar{\tau}) d X^{\mu}(\bar{\sigma},\bar{\tau}) d X^{\mu'}(\bar{\sigma}',\bar{\tau}). \label{LineElement}
\end{align}
The inverse metric $\mathbf{G}^{IJ}(\bar{h}, X_{\hat{D}_{T}}(\bar{\tau}), \bar{\tau})$ is defined by $\mathbf{G}_{IJ}\mathbf{G}^{JK}=\mathbf{G}^{KJ}\mathbf{G}^{JI}=\delta^{K}_{I}$, where $\delta^{d}_{d}=1$ and $\delta^{(\mu' \bar{\sigma}')}_{(\mu \bar{\sigma})}=\frac{1}{\bar{e}(\bar{\sigma},\bar{\tau})} \delta^{\mu'}_{\mu} \delta(\bar{\sigma} - \bar{\sigma}')$.

\section{String background configuration}
In this section, we define configurations of fields that include the string backgrounds. Second, we show that the equations of motion where the configurations are substituted are equivalent to the equations of motions of the string backgrounds.

From eq.~(\ref{action of bos string-geometric model}), the equations of motion of $\mathbf{G}_{IJ}$, $\Phi$, and $\mathbf{B}_{IJ}$ are given by 
\begin{align}
\label{eom bos string-geometric model 3.1}
&\mathbf{R}_{IJ}-\frac{1}{4} \mathbf{H}_{IL_{1} L_{2}} \mathbf{H}_{J}^{L_{1}L_{2}} + 2 \nabla_{I} \nabla_{J} \Phi 
-\frac{1}{2}\mathbf{G}_{IJ}
(\mathbf{R}-4 \nabla_{L_1} \Phi \nabla^{L_1} \Phi + 4 \nabla_{L_1} \nabla^{L_1} \Phi - \frac{1}{2} | \mathbf{H} |^{2})
 = 0,\\
\label{eom bos string-geometric model 3.2}
&\mathbf{R}-4 \nabla_{L_1} \Phi \nabla^{L_1} \Phi + 4 \nabla_{L_1} \nabla^{L_1} \Phi - \frac{1}{2} | \mathbf{H} |^{2}  = 0, \\ 
\label{eom bos string-geometric model 3.3}
&\nabla_{L_1} \left(  e^{-2 \Phi} \mathbf{H}^{L_1 I J} \right) = 0,
\end{align}
respectively.

We consider the following ansatz, which we call the string background configuration, 
\begin{description}
\item{Metric:}\footnotemark 
\begin{align}
&\mathbf{G}_{dd} \left( X^{d}_{\hat{D}_{T}}, X_{\hat{D}_{T}}\right) = -1, \label{Gdd} \\ 
&\mathbf{G}_{(\mu_{1} \bar{\sigma}_{1})(\mu_{2} \bar{\sigma}_{2} )} \left( X^{d}_{\hat{D}_{T}}, X_{\hat{D}_{T}}\right)= \beta (\bar{\sigma}_{1}) \delta_{\bar{\sigma}_{1} \bar{\sigma}_{1} } G_{\mu_{1} \mu_{2}} \left( X_{\hat{D}_{T}}(\bar{\sigma}_{1}) \right) \delta_{\bar{\sigma}_{1} \bar{\sigma}_{2} }, \\
&\text{the others} = 0,
\end{align}
\footnotetext{The inverse is obtained as 
\begin{align*}
G ^{(\mu_{1} \bar{\sigma}_{1} )(\mu_{2} \bar{\sigma}_{2})} \left( X^{d}_{\hat{D}_{T}}, X_{\hat{D}_{T}}\right) = \beta^{-1} (\bar{\sigma}_{1}) \delta_{\bar{\sigma}_{1} \bar{\sigma}_{1} } ^{-1} \delta_{\bar{\sigma}_{1} \bar{\sigma}_{2} } G^{\mu_{1} \mu_{2}} \left( X_{\hat{D}_{T}}(\bar{\sigma}_{1}) \right).
\end{align*}}
\item{Scalar field:}
\begin{align}
\Phi \left( X^{d}_{\hat{D}_{T}}, X_{\hat{D}_{T}}\right) = \int  d \bar{\sigma} \bar{e}(\bar{\sigma},\bar{\tau})  f(\bar{\sigma}) \delta_{\bar{\sigma} \bar{\sigma}} \phi \left( X_{\hat{D}_{T}}(\bar{\sigma}) \right),
\end{align}
\item{2-form field:}
\begin{align}
&\mathbf{B}_{(\mu_{1} \bar{\sigma}_{1})(\mu_{2} \bar{\sigma}_{2}) } \left( X^{d}_{\hat{D}_{T}}, X_{\hat{D}_{T}}\right) = b (\bar{\sigma}_{1}) \delta_{\bar{\sigma}_{1} \bar{\sigma}_{1} } B_{\mu_{1} \mu_{2}} \left( X_{\hat{D}_{T}}(\bar{\sigma}_{1}) \right) \delta_{\bar{\sigma}_{1} \bar{\sigma}_{2}  }, \\
&\text{the others} = 0, \label{others}
\end{align}
\end{description}
where $\delta_{\bar{\sigma}_{1} \bar{\sigma}_{2}}=\frac{1}{\bar{e}(\bar{\sigma},\bar{\tau})}\delta(\bar{\sigma}_{1} - \bar{\sigma}_{2})$, $G_{\mu_{1} \mu_{2}} \left( x \right)$ is a symmetric tensor field, $\phi \left( x \right)$ is a scalar field and $B_{\mu_{1} \mu_{2}} \left( x \right)$ is an anti-symmetric tensor field on a 26-dimensional spacetime. 

We remark that the string background configuration has a non-trivial dependence on the worldsheet. The consistent truncation will be ensured due to the relation between the worldsheet dependence of the fields and of the indices of the string geometry fields. For example, see $\bar{\sigma}_1$ dependence on the string background configuration for the metric. In addition, the factor $\delta_{\bar{\sigma} \bar{\sigma} }$ reflects that the point particle limit is a field theory. 

By substituting the string background configuration,
the left hand side of the Einstein equation becomes
\begin{eqnarray}
&&\bold{R}_{(\mu \bar{\sigma}_1) (\nu \bar{\sigma}_2)} 
-\frac{1}{2}\bold{G}_{(\mu \bar{\sigma}_1) (\nu \bar{\sigma}_2)} \bold{R}
\nonumber \\
&=&
\delta_{\bar{\sigma}_1 \bar{\sigma}_2}
\delta_{\bar{\sigma}_1 \bar{\sigma}_1}
\left(R_{\mu \nu} \left( X_{\hat{D}_T}(\bar{\sigma}_{1}) \right)
-\frac{1}{2}
G_{\mu \nu} \left( X_{\hat{D}_T}(\bar{\sigma}_{1}) \right)
\int d \bar{\sigma} \bar{e}(\bar{\sigma})   \delta_{\bar{\sigma} \bar{\sigma}} 
R \left( X_{\hat{D}_T}(\bar{\sigma}) \right)  \right).
\end{eqnarray}
As one can see in this formula, if an equation of motion includes a trace (in $\bold{R}$ in this case), the reduced equation of motion includes an extra summation $\int d \bar{\sigma} \bar{e}(\bar{\sigma})   \delta_{\bar{\sigma} \bar{\sigma}}$ against the equation of motion of the string backgrounds. Fortunately, eq.~(\ref{eom bos string-geometric model 3.2}) is proportional to the terms including the trace in eq.~(\ref{eom bos string-geometric model 3.1}). Then, under eq.~(\ref{eom bos string-geometric model 3.2}), eq.~(\ref{eom bos string-geometric model 3.1}) becomes
\begin{align}
\label{eom2 bos string-geometric model 3.1}
\mathbf{R}_{IJ}-\frac{1}{4} \mathbf{H}_{IL_{1} L_{2}} \mathbf{H}_{J}^{L_{1}L_{2}} + 2 \nabla_{I} \nabla_{J} \Phi 
 = 0,
\end{align}
which does not include the trace. This is a result of our choice of the action ~(\ref{action of bos string-geometric model}).

Consequently, eqs.~(\ref{eom2 bos string-geometric model 3.1}), (\ref{eom bos string-geometric model 3.2}) and  (\ref{eom bos string-geometric model 3.3}) become
\begin{align}
\label{eom bos string-geometric model 3.1.1}
&\left[ R_{\mu \nu} \left( X_{\hat{D}_{T}} (\bar{\sigma}_{1}) \right) - \frac{1}{4} \beta^{-2}( \bar{\sigma}_{1}) b^{2}(\bar{\sigma}_{1}) H_{\mu \mu_{1} \mu_{2}} \left( X_{\hat{D}_{T}} (\bar{\sigma}_{1}) \right) H_{\nu}^{\mu_{1} \mu_{2}} \left( X_{\hat{D}_{T}} (\bar{\sigma}_{1}) \right) + 2f(\bar{\sigma}_{1}) \nabla_{\mu } \nabla_{ \nu } \phi \left( X_{\hat{D}_{T}} (\bar{\sigma}_{1}) \right) \right] \notag \\
&\times \delta_{\bar{\sigma}_{1} \bar{\sigma}_{2}} \delta_{\bar{\sigma}_{1} \bar{\sigma}_{1}} = 0,\\
\label{eom bos string-geometric model 3.2.2} 
&\int d \bar{\sigma} \bar{e}(\bar{\sigma},\bar{\tau})   \delta_{\bar{\sigma} \bar{\sigma}} \beta^{-1}(\bar{\sigma}) \left[ R \left( X_{\hat{D}_{T}} (\bar{\sigma}) \right) -4  f^{2} (\bar{\sigma}) \partial_{\mu_{1} } \phi \left( X_{\hat{D}_{T}} (\bar{\sigma}) \right) \partial^{\mu_{1} } \phi \left( X_{\hat{D}_{T}} (\bar{\sigma}) \right) \right. \notag \\
&\hspace{4cm}\left.   + 4   f (\bar{\sigma}) \nabla_{\mu_{1} }  \nabla^{\mu_{1} }  \phi \left( X_{\hat{D}_{T}} (\bar{\sigma}) \right) - \frac{1}{2} \beta^{-2} (\bar{\sigma}) b^{2} (\bar{\sigma}) |H \left( X_{\hat{D}_{T}} (\bar{\sigma}) \right) |^{2}  \right] =0,
\end{align}
\begin{align}
\label{eom bos string-geometric model 3.3.3}
&\Big[ -2 f (\bar{\sigma}_{1}) \partial_{\mu_{1} }  \phi \left( X_{\hat{D}_{T}} (\bar{\sigma}_{1}) \right) H^{\mu_{1} \mu \nu}  \left( X (\bar{\sigma}_{1}) \right) + \nabla_{\mu_{1} }  H^{\mu_{1} \mu \nu} \left( X_{\hat{D}_{T}} (\bar{\sigma}_{1}) \right)  \Big] \times \delta_{\bar{\sigma}_{1} \bar{\sigma}_{2}} \delta_{\bar{\sigma}_{1} \bar{\sigma}_{1}}  = 0,
\end{align}
respectively, and the other components are automatically satisfied. We use the Einstein notation for only the index $\mu$.

Eqs.~(\ref{eom bos string-geometric model 3.1.1})~$\sim$~(\ref{eom bos string-geometric model 3.3.3}) are satisfied if and only if
\begin{align*}
f(\bar{\sigma})&=1, \qquad \beta^{-2}(\bar{\sigma}) b^{2}(\bar{\sigma})=1.
\end{align*}
and
$G_{\mu_1 \mu_2}$, $\phi$ and $B_{\mu_1 \mu_2}$ satisfy 
the equations of motion of the string backgrounds 
\begin{align}
\label{eom bos SUGRA1}
&R_{\mu_{1} \mu_{2} }-\frac{1}{4} H_{\mu_{1} \nu_{1} \nu_{2} } H_{\mu_{2} }^{\nu_{1} \nu_{2} } + 2 \nabla_{\mu_{1} } \nabla_{\mu_{2} } \phi
-\frac{1}{2}G_{\mu_{1} \mu_{2} }
(R-4 \nabla_{\nu_{1} } \phi \nabla^{\nu_{1} } \phi + 4 \nabla_{\nu_{1} } \nabla^{\nu_{1} } \phi - \frac{1}{2} |H|^{2})= 0, \\
\label{eom bos SUGRA2}
&R-4 \nabla_{\mu_{1} } \phi \nabla^{\mu_{1} } \phi + 4 \nabla_{\mu_{1} } \nabla^{\mu_{1} } \phi - \frac{1}{2} |H|^{2}  = 0, \\ 
\label{eom bos SUGRA3}
&\nabla_{\nu_{1} } \left(  e^{-2 \phi} H^{\nu_{1}  \mu_{1} \mu_{2}} \right) = 0,
\end{align}
which are derived from the action,
\begin{align}
\label{action of bos sugra}
&S = \frac{1}{2 \kappa^{2}_{10}}  \int d^{10}x \sqrt{G}  e^{ -2 \phi } \left[ R + 4 \nabla_{\mu} \phi \nabla^{\mu} \phi - \frac{1}{2} |H|^{2}  \right],
\end{align}
where 
$|H|^{2}:= \frac{1}{3!} G^{\mu_{1} \nu_{1}} G^{\mu_{2} \nu_{2}} G^{\mu_{3} \nu_{3}} H_{\mu_{1} \mu_{2} \mu_{3}} H_{\nu_{1} \nu_{2} \nu_{3}}$.
Therefore, we conclude that the string backgrounds can be embedded into the string geometry model in the sense of the consistent truncation.

\section{Equations that determine a string background}
Because the string background configuration eqs.~(\ref{Gdd})~$\sim$~(\ref{others}) is stationary with respect to the string geometry time $\bar{\tau}$, the energy of it is defined as
\begin{eqnarray}
E&=&\int \mathcal{D} h  \mathcal{D} X T_{00}
\nonumber \\
&=&\int \mathcal{D} h  \mathcal{D} X
\left(
-2 \nabla_{0} \nabla_{0} \Phi 
+\frac{1}{4} \mathbf{H}_{0L_{1} L_{2}} \mathbf{H}_{0}^{L_{1}L_{2}}+G_{00}(-2 \nabla_{I} \Phi \nabla^{I} \Phi + 2 \nabla_{I} \nabla^{I} \Phi - \frac{1}{4} | \mathbf{H} |^{2})
\right) \nonumber \\
&=&\int \mathcal{D} X \int \mathcal{D} e \int d \bar{\sigma} \bar{e}(\bar{\sigma})   \delta_{\bar{\sigma} \bar{\sigma}} 
(2 \nabla_{\mu} \phi \left( X(\bar{\sigma}) \right) \nabla^{\mu} \phi  - 2 \nabla_{\mu} \nabla^{\mu} \phi + \frac{1}{4} | H |^{2})  \nonumber \\
&=&\int d^{10}x \sqrt{-G(x)}
(2 \nabla_{\mu} \phi(x)\nabla^{\mu} \phi  - 2 \nabla_{\mu} \nabla^{\mu} \phi + \frac{1}{4} | H |^{2}).  \label{energy}
\end{eqnarray}
On the second line in the above formula, we have substituted eqs.~(\ref{Gdd})~$\sim$~(\ref{others}) and obtained the third line. On the third line, we have reguralized the integral over the embedding function as
\begin{eqnarray}
&&\int \mathcal{D} X=\prod_{j=1}^N \int d^{10}x_j \sqrt{-G(x_j)}
\nonumber \\
&&\int d^{10}x_j \sqrt{-G(x_j)}=1
\nonumber \\
&&\int \mathcal{D} e \int d \bar{\sigma} \bar{e}(\bar{\sigma})   \delta_{\bar{\sigma} \bar{\sigma}} = \frac{1}{N}\sum_{i=1}^N,
\end{eqnarray}
and obtained the fourth line.

Therefore, in string geometry theory, a string background is determined by minimizing the energy (\ref{energy}) of the solutions to the equations of motions: eqs.~(\ref{eom bos SUGRA1}), ~(\ref{eom bos SUGRA2}) and ~(\ref{eom bos SUGRA3}). In other words, by using the method of Lagrange multiplier, the equations that determine string backgrounds are obtained by differentiating 
\begin{eqnarray}
\tilde{E}
=
E
+\int d^{10}x \sqrt{-G(x)}
(\lambda_G^{\mu \nu}(x) f^G_{\mu \nu}(x)
+\lambda_{\phi}(x) f^{\phi}(x)
+\lambda_B^{\mu \nu}(x) f^B_{\mu \nu}(x))
\nonumber
\end{eqnarray}
with respect to 
the string backgrounds $G_{\mu \nu}(x)$, $\phi(x)$, and $B_{\mu \nu}(x)$
and
the Lagrange multipliers $\lambda_G^{\mu \nu}(x)$, $\lambda_{\phi}(x)$ and $\lambda_B^{\mu \nu}(x)$, 
where $f^G_{\mu \nu}(x)=0$, $f^{\phi}(x)=0$ and $f^B_{\mu \nu}(x)=0$ represent the equations of motions: eqs.~(\ref{eom bos SUGRA1}), ~(\ref{eom bos SUGRA2}) and ~(\ref{eom bos SUGRA3}), respectively.

\section{Conclusion and Discussion}
In this paper, we have shown that arbitrary configurations of the string backgrounds are embedded in configurations of fields of a string geometry model as in eqs.~(\ref{Gdd})~$\sim$~(\ref{others}). Especially, we have shown that the equations of motion where the configurations are substituted are equivalent to the equations of motions of the string backgrounds. This means that classical dynamics of the string backgrounds are described as a part of classical dynamics in string geometry theory. This fact supports the conjecture that string geometry theory is a non-perturbative formulation of string theory.


Furthermore, we define an energy of the string background configurations, because they are stationary with respect to the string geometry time $\bar{\tau}$. Thus, a string background can be determined by minimizing the energy of the solutions to the equations of motions of the string backgrounds. Therefore, we conclude that string geometry theory includes a non-perturbative effect that determines a string background.

Here we discuss future directions. It is important to derive the path-integral of the nonlinear sigma model~\cite{Callan:1985ia,Fradkin:1985ys,Polchinski:1998rr} from fluctuations around the fixed background eqs.~(\ref{Gdd})~$\sim$~(\ref{others}). We already showed that this is true when the string background is flat \cite{Sato:2017qhj, Sato:2020szq}. The string background fields in the string background configuration eqs.~(\ref{Gdd})~$\sim$~(\ref{others}) depend not only on the string zero modes $x^{\mu}$ but also on the other modes of $X^{\mu}_{\hat{D}_{T}}(\bar{\sigma})$. In addition, the consistent truncation of the equations of motion is valid without taking $\alpha' \to 0$ limit, which corresponds to $X (\bar{\sigma}) \to x$. This fact will be important to derive the non-linear sigma model since the string backgrounds in the non-linear sigma model depend not only on the string zero modes $x$ but also on the other modes of $X(\bar{\sigma})$. We are also interested in the supersymmetric case. With a little alteration, due to the existence of the RR fields, the gauge field and the Grassmann coordinates, we may show that the consistent truncation is valid in the supersymmetric case~\cite{Honda-Sato}.
 

\section*{Acknowledgements}
We would like to thank  
K. Hashimoto,
Y. Hyakutake,
T. Onogi,
S. Sugimoto,
Y. Sugimoto,
S. Yamaguchi,
and especially 
H. Kawai and A. Tsuchiya
for long and valuable discussions.


\bibliographystyle{prsty}

\end{document}